\documentclass[twocolumn,aps,pra,showpacs,floatfix]{revtex4-1}

\usepackage{latexsym}
\usepackage{amsmath}
\usepackage{amssymb}
\usepackage{latexsym}
\usepackage{xcolor}
\usepackage{graphicx}
\usepackage{cancel}
\usepackage{bbm}
\usepackage{bm}
\usepackage{maybemath}

\newcommand{\dd}{{\mathrm d}}
\newcommand{\ii}{{\mathrm i}}
\newcommand{\ee}{{\mathrm e}}

\newcommand{\bfH}{\mathbf{H}}

\newcommand{\calE}{\mathcal{E}}
\newcommand{\calF}{\mathcal{F}}
\newcommand{\calH}{\mathcal{H}}
\newcommand{\calO}{\mathcal{O}}
\newcommand{\calP}{\mathcal{P}}
\newcommand{\calT}{\mathcal{T}}

\newcommand{\alphap}{\vec\alpha\cdot\vec p}

\newcommand{\rmW}{\mathrm{W}}
\newcommand{\rmDS}{\mathrm{DS}}
\newcommand{\rmFD}{\mathrm{FD}}
\newcommand{\rmTD}{\mathrm{TD}}
\newcommand{\rmTG}{\mathrm{TG}}

\newcommand{\sdp}{\vec\sigma\cdot\vec p}

\newcommand{\rr}{\frac{r_s}r}
\newcommand{\rt}{\frac{r_s}{2r}}
\newcommand{\rs}{\frac{r_s^{\,2}}{r^2}}
\newcommand{\pp}{|\vec p|}

\begin{document}

\title{Ultrarelativistic Decoupling Transformation for 
Generalized Dirac Equations}

\author{J. H. Noble}
\affiliation{Department of Physics,
Missouri University of Science and Technology,
Rolla, Missouri 65409, USA}

\author{U. D. Jentschura}
\email{ulj@mst.edu}
\affiliation{Department of Physics,
Missouri University of Science and Technology,
Rolla, Missouri 65409, USA}

\begin{abstract}
The Foldy--Wouthuysen transformation is known to uncover the nonrelativistic
limit of a generalized Dirac Hamiltonian, lending an intuitive physical
interpretation to the effective operators within
Schr\"{o}dinger--Pauli theory.  We here discuss the opposite, ultrarelativistic
limit which requires the use of a fundamentally different expansion 
where the leading kinetic term in the Dirac equation
is perturbed by the mass of the particle and
other interaction (potential) terms, rather than vice versa.  The
ultrarelativistic decoupling transformation is applied to
free Dirac particles (in the Weyl basis) and to 
high-energy tachyons, which are faster-than-light particles described by a
fully Lorentz-covariant equation. The effective gravitational interactions
are found. For tachyons, the dominant gravitational interaction
term in the high-energy limit is shown to be attractive,
and equal to the leading term for subluminal Dirac particles
(tardyons) in the high-energy limit. 
\end{abstract}


\maketitle

%
%
\section{Introduction}
\label{sec1}

The Foldy--Wouthuysen transformation~\cite{FoWu1950} is an established method,
used to calculate the nonrelativistic limit of effective Hamiltonians
describing spin-$1/2$ particles.  The procedure has been applied with good
effect to the Dirac--Coulomb Hamiltonian~\cite{BjDr1964,ItZu1980}, uncovering
the fine-structure terms for atomic levels, notably, the zitterbewegung term,
and the Russell--Saunders (spin-orbit) coupling (Thomas precession).  Recently,
the analogue of the Russell--Saunders coupling in a gravitational field (the
Fokker precession, see Refs.~\cite{Je2013,JeNo2013pra,JeNo2014jpa}) has been
recovered from the gravitationally coupled Dirac Hamiltonian, which is referred
to as the Dirac--Schwarzschild Hamiltonian~\cite{JeNo2013pra}.

The Foldy--Wouthuysen program, in its original form~\cite{FoWu1950},
is inherently perturbative in nature. In
a typical case, the structure of a generalized Dirac Hamiltonian is given as
$\vec\alpha \cdot \vec p + \beta \, m + \delta H$ (in the standard Dirac
representation of the Clifford algebra, see Appendix~\ref{appa}). Here, the
``dominant'' term is taken as $\beta \, m$, where $\beta$ is the $4\times 4$
Dirac $\beta$ matrix, $\vec\alpha \cdot \vec p$ is the kinetic operator
($\vec\alpha$ is the vector of Dirac $\alpha$ matrices, and $\vec p$ is the
momentum operator), and $\delta H$ contains the potential terms.  One then
expands about a Dirac particle ``at rest'', with the dominant term given by the
``rest mass'' term $\beta m$.  The Foldy--Wouthuysen procedure then uncovers
the leading nonrelativistic kinetic term $\vec\alpha \cdot \vec p \to \vec
p^{\,2}/(2m) + \dots$ and transforms the potential terms $\delta H$ into a form
where the operators acquire an intuitive physical interpretation.  At some risk
to oversimplification, one can say that the Foldy--Wouthuysen transformation
applies to the regime $| \vec\alpha \cdot \vec p | \ll | \beta m |$, and 
$| \delta H | \ll | \beta m |$.

In some cases, such as for a free Dirac particle~\cite{BjDr1964}, it is
sometimes possible to perform the transformation without any perturbative
expansion in the momenta or other expansion parameters.  There have been
attempts to generalize the idea of a nonperturbative method to more general
Hamiltonians, and a set of interesting identities have been derived 
Ref.~\cite{ErKo1958}. However, the alternative Foldy--Wouthuysen
transformation~\cite{ErKo1958} 
suffers from an explicit breaking of the parity symmetry in the
transformation, which involves the fifth current, and is known to produce
spurious parity-breaking terms in a number of applications, e.g., to the
Dirac--Coulomb Hamiltonian (for an overview, 
see Refs.~\cite{GoNe2014,Je2014dirac,JeNo2014jpa}).  In
general, nonperturbative methods (in the momenta of the Dirac particles) can
only be applied when considerable additional information is available for a
specific Hamiltonian under investigation, and when additional approximations
are made, such as the neglecting terms of second order and higher in the field
strengths [see Eq.~(21) of Ref.~\cite{Si2008pra}].

To the best of our knowledge, the opposite perturbative expansion, namely,
perturbation theory of a Dirac Hamiltonian about the ultrarelativistic limit,
has not yet been considered in the literature; it is the subject of the current
paper.  This expansion has to follow a fundamentally different paradigm; in the
ultrarelativistic limit, mass terms and potential terms are suppressed in
comparison to the kinetic term; the expansion is valid in the regime 
$| \beta m |\ll | \vec\alpha \cdot \vec p |$, and 
$| \delta H | \ll | \vec\alpha \cdot \vec p |$.
Ultrarelativistic particles are best described in the helicity basis [Chap.~23
of Ref.~\cite{BeLiPi1982vol4}], while in fact, the solutions of the free Dirac
equation approach those of the Weyl equation in the massless limit (see
Chap.~2.4.3 on p.~87 of Ref.~\cite{ItZu1980}).  The Weyl equation describes
massless spin-$1/2$ particles, which transform under the fundamental
$(\tfrac12,0)$ representation of the Lorentz group and travel exactly at the
speed of light (these are the ``neutrinos in the original standard model'').

We here investigate the ultrarelativistic decoupling transformation with a
special emphasis on the gravitational coupling of a particle to a central
gravitational field.  To this end, in Sec.~\ref{sec2}, we briefly recall the
underlying covariant formalism, distinguishing the case of a ``normal''
(subluminal) Dirac particle from a particle described by the tachyonic Dirac
equation~\cite{Je2013,JeNo2013pra,JeNo2014jpa}. The latter equation describes
faster-than-light particles, still in a fully Lorentz-covariant 
formalism~\cite{ChHaKo1985}. The ultrarelativistic limit
specifically is relevant to tachyons because these particles
cannot travel slower than light; they remain superluminal upon Lorentz
transformation~\cite{BiDeSu1962,Fe1967,ArSu1968,BiSu1969,DhSu1968}. 
In the ultrarelativistic limit,
the particle's speed approaches the light cone and the influence of
tardyonic as well as tachyonic mass terms are suppressed in comparison to the
kinetic terms.  In Sec.~\ref{sec3}, the ultrarelativistic decoupling
transformation is applied to gravitationally coupled tardyonic and 
tachyonic particles.  Conclusions are reserved for Sec.~\ref{sec4}.  

%
%
\section{Free Particles}
\label{sec2}

%
%
\subsection{Free Tardyonic Transformation}
\label{sec2A}

In principle it is well known that the Weyl equation, 
which describes a massless spin-$1/2$ particle, splits into 
two equations, describing a left-handed and a right-handed spinor
(see Chap.~23 of Ref.~\cite{BeLiPi1982vol4} and p.~87 of Ref.~\cite{ItZu1980}),
\begin{align}
\ii \, \partial_t \psi_L =& \; H_L \, \psi_L \,,
\qquad
\qquad
H_L = -\vec\sigma \cdot \vec p \,,
\\[0.133ex]
\ii \, \partial_t \psi_R =& \; H_R \, \psi_R \,,
\qquad
\qquad
H_R = \vec\sigma \cdot \vec p \,,
\end{align}
The Weyl equations break parity; a left-handed spinor 
transforms into a right-handed solution under the parity 
operation.  However, it is well known that the Dirac 
equation, whose bispinor solutions are constructed
by stacking the helicity spinors on top of each other, 
conserves parity~\cite{Sr2007}.

The massless equation, in turn, corresponds to the 
ultrarelativistic limit for a massive Dirac particle;
we would thus expect that the Dirac equation splits into 
two equations, describing left- and right-handed Weyl spinors,
in this limit.
Thus, if we are to recover the massless (Weyl) limit, plus corrections,
from the Dirac equation, then we need to necessarily invoke a 
parity-breaking transformation.
We start from the free Dirac Hamiltonian
\begin{equation}
\label{HFD}
H_\rmFD = \vec \alpha \cdot \vec p + \beta \, m 
= \left( \begin{array}{cc}
m & \vec\sigma \cdot \vec p \\
\vec\sigma \cdot \vec p & - m 
\end{array} \right)
\end{equation}
and invoke the following unitary, parity-breaking transformation
\begin{equation}
\label{defU}
U = \frac{1}{\sqrt{2}} \, (1 - \beta \, \gamma^5) \,,
\qquad
U^{-1} = U^{\rm T} = \frac{1}{\sqrt{2}} \, (1 + \beta \, \gamma^5) \,,
\end{equation}
which transforms $H_\rmFD$ into 
$\bfH_\rmFD = U \, H_\rmFD \, U^{-1}$,
\begin{equation}
\bfH_\rmFD = -\beta \vec \Sigma \cdot \vec p + \gamma^5 \, m = 
\left( \begin{array}{cc} 
-\vec\sigma \cdot \vec p & m \\
m & \vec\sigma \cdot \vec p 
\end{array} \right)
\end{equation}
The initial rotation with the $U$ matrix corresponds
to a change of the basis of the Dirac algebra, into the 
so-called Weyl basis (see Appendix~\ref{appa}).
The mass terms are now off-diagonal and we may try to 
eliminate them by an ultrarelativistic 
decoupling (ultrarelativistic Foldy--Wouthuysen) transformation.
To this end we define the energy operator
\begin{equation}
\label{defcalE}
\calE = -\vec \Sigma \cdot \vec p \,,
\end{equation}
and the transformation
(see Sec~4.2 of Ref.~\cite{BjDr1964})
and Sec.~2.2.4 of Ref.~\cite{ItZu1980})
\begin{equation}
S_\rmFD = - \ii \, \beta \, \gamma^5 \, \frac{m}{\calE} \Theta \,,
\qquad
S^+_\rmFD = S_\rmFD \,.
\end{equation}
so that the unitary transformation $U_\rmFD$ becomes
\begin{equation}
U_\rmFD = \exp( \ii \, S_\rmFD ) = 
\cos\left( \frac{m \, \Theta}{| \vec p|} \right) + 
\beta \, \gamma^5 \, \frac{| \vec p |}{\calE}
\sin\left( \frac{m \, \Theta}{|\vec p|} \right) 
\end{equation}
Choosing $\Theta$ so that 
\begin{subequations}
\label{Theta1}
\begin{align}
\cos\left( 2 \frac{m \, \Theta}{| \vec p|} \right) = & \;
\frac{|\vec p|}{\sqrt{\vec p^{\,2} + m^2}} \,,
\\[0.133ex]
\sin\left( 2 \frac{m \, \Theta}{| \vec p|} \right) = & \;
\frac{m}{\sqrt{\vec p^{\,2} + m^2}} \,,
\end{align}
\end{subequations}
one finally obtains
\begin{equation}
\label{calHFD}
\calH_{\rm FD} = 
U_\rmFD \, \bfH_{\rm FD}  \, U_\rmFD^{-1}  =
\frac{\calE}{| \vec p |} \, \sqrt{\vec p^{\,2} + m^2 } \,.
\end{equation}
In explicit $(2 \times 2)$-matrix subform,
\begin{align}
\calH_{\rm FD} =& \;
- \beta \, \frac{\vec \Sigma \cdot \vec p}{| \vec p |} \,
\sqrt{\vec p^{\,2} + m^2 }
\nonumber\\[2.133ex]
=& \; \left( \begin{array}{cc}
-\dfrac{\vec\sigma \cdot \vec p}{| \vec p |} \,
\sqrt{\vec p^{\,2} + m^2 } & 0 \\
0 & \dfrac{\vec\sigma \cdot \vec p}{| \vec p |} \, 
\sqrt{\vec p^{\,2} + m^2 } 
\end{array} \right)\,,
\end{align}
it becomes clear that the separation into a 
left-handed (upper diagonal) and a right-handed (lower diagonal)
Hamiltonian has been achieved.

The eigenstates of the Hamiltonian~\eqref{calHFD} fulfill 
\begin{subequations}
\label{eigenS}
\begin{equation}
\ii \, \partial_t \psi_i(t, \vec r) = 
  \calH_\rmFD \, \psi_i(t, \vec r) \,,
\qquad i=1,2,3,4\,.
\end{equation}
The first two solutions can be written as
\begin{align}
\psi_1(t, \vec r) =& \;
\left( \begin{array}{c} a_-(\vec k) \\ 0 \end{array} \right) \, 
\ee^{-\ii E \, t + \ii \, \vec k \cdot \vec r} \,,
\qquad
H_\rmFD \, \psi_1 = E \, \psi_1 \,,
\\[2ex]
\psi_2(t, \vec r) =& \;
\left( \begin{array}{c} a_-(\vec k) \\ 0 \end{array} \right) \,
\ee^{\ii E \, t - \ii \, \vec k \cdot \vec r} \,,
\qquad
H_\rmFD \, \psi_2 = -E \, \psi_2 \,.
\end{align}
The physical momentum is $\vec k$, and the helicity 
eigenvalue is negative for both solutions,
$\vec\Sigma \cdot \hat k \, \psi_{1,2} = -\psi_{1,2}$,
and $E = \sqrt{\vec k^2 + m^2}$ 
(here, $\hat k$ is the unit vector in the $\vec k$
direction). 
The solution $\psi_2$ describes an antiparticle.
The two solutions of right-handed helicity are
\begin{align}
\psi_3(t, \vec r) =& \;
\left( \begin{array}{c} 0 \\ a_+(\vec k) \end{array} \right) \,
\ee^{-\ii E \, t + \ii \, \vec k \cdot \vec r} \,,
\qquad
H_\rmFD \, \psi_3 = E \, \psi_3 \,,
\\[2ex]
\psi_4(t, \vec r) =& \;
\left( \begin{array}{c} 0 \\ a_+(\vec k) \end{array} \right) \,
\ee^{\ii E \, t - \ii \, \vec k \cdot \vec r} \,,
\qquad
H_\rmFD \, \psi_4 = -E \, \psi_4 \,,
\end{align}
\end{subequations}
The helicity is positive for these two solutions,
$\vec\Sigma \cdot \hat k \, \psi_{3,4} = \psi_{3,4}$,
with $\psi_4$ describing an antiparticle.
The eigenvalues of the $\calE$
operator for $\psi_{1,2,3,4}$ are 
$E,-E,-E,E$, respectively.
If we apply the formalism to a Dirac neutrino,
them $\psi_1$ would describe a left-handed neutrino,
$\psi_2$ would describes a left-handed antineutrino,
whereas $\psi_3$ and $\psi_4$ would describe 
right-handed neutrinos and right-handed antineutrinos,
respectively. For completeness, we recall the form of the 
helicity spinors~\cite{ItZu1980},
\begin{subequations}
\begin{align}
a_+(\vec k) =& \; \left( \begin{array}{c}
\cos\left(\frac{\theta}{2}\right) \\[0.33ex]
\sin\left(\frac{\theta}{2}\right) \, \ee^{\ii \, \varphi} \\
\end{array} \right) \,,
\\[0.133ex]
a_-(\vec k) =& \; \left( \begin{array}{c}
-\sin\left(\frac{\theta}{2}\right) \, \ee^{-\ii \, \varphi} \\[0.33ex]
\cos\left(\frac{\theta}{2}\right) \\
\end{array} \right) \,.
\end{align}
\end{subequations}
where $\theta$ and $\varphi$ are the polar and azimuthal 
angles of the $\vec k$ vector.

%
%
\subsection{Free Tachyonic Transformation}
\label{sec2B}

As we have just shown, one may accomplish an exact diagonalization (in spinor
space) of the free Dirac Hamiltonian using the ultrarelativistic decoupling
transformation.  However, one might counter argue that this result is in
principle familiar: An exact diagonalization can also be accomplished using the
Foldy--Wouthuysen transformation (in its original form) for the free Dirac
Hamiltonian (see Sec.~4.2 of Ref~\cite{BjDr1964}). The ultrarelativistic
transform leads to a form which asymptotically is equal to the Weyl Hamiltonian
(helicity basis) of a massless particle, as it should be (in the ultrarelativistic
limit). Here, we shall make the point that, unlike the original
Foldy--Wouthuysen transform, which can only be applied to tardyons, the
ultrarelativistic decoupling can also be used for Lorentz-invariant
tachyons~\cite{ChHaKo1985,JeWu2012epjc,JeWu2012jpa}, whose velocity remains
superluminal upon Lorentz
transformation~\cite{BiDeSu1962,Fe1967,ArSu1968,BiSu1969,DhSu1968}.

A few general remarks on tachyonic spin-$1/2$ particles might be in order.  The
tachyonic neutrino hypothesis remains one of the driving forces behind the
study of the tachyonic Dirac equation~\cite{ChHaKo1985}.  The 
algbraic structures underlying the tachyonic
spin-$1/2$ equation have recently been studied in greater 
depth (see Refs.~\cite{Ci1998,Ch2000,Ch2001,JeWu2012jpa,JeWu2014}
and references therein).  
Pertinent potentially relevant
astrophysical observations have recently been
recorded in Refs.~\cite{Eh2000,Eh2012,Eh2013,Eh2015}; other theoretical studies
concern Dirac equations with Lorentz-violating terms which can lead to
superluminal propagation for neutrinos~\cite{KoMe2012,JeEtAl2014}. 
The tachyonic Dirac Hamiltonian has recently been identified
as a pseudo-Hermitian (``$\gamma^5$--Hermitian'') Hamiltonian
in Ref.~\cite{JeWu2012jpa}.
Independent of the phenomenological relevance of the concept of tachyons,
the current section of our paper, and Sec.~\ref{sec3B}
demonstrate that it is possible to uncover the leading terms of generalized
pseudo-Hermitian~\cite{Pa1943,BeBo1998,Mo2002i,Mo2002ii,Mo2002iii} 
Dirac Hamiltonians in the ultrarelativistic limit using the 
relativistic decoupling transformation.

The accepted generalized Dirac Hamiltonian for a free tachyonic 
Dirac particle is given as~\cite{ChHaKo1985,JeWu2012epjc,JeWu2012jpa}
\begin{equation}
H_{\rmTD}= \alphap + \beta\,\gamma^5\,m\,,
\end{equation}
which is $\gamma^5$-Hermitian,
$H_\rmTD = \gamma^5 \, H^+_\rmTD \, \gamma^5$.
We then follow the same procedure outlined in Sec.~\ref{sec2A}, and begin by
performing the initial rotation $U$ [see Eq.~\eqref{defU}], giving us
\begin{equation}
\bfH_\rmTD = U \, H_\rmTD \, U^{-1}
= \beta\,\calE + \beta\,\gamma^5\,m \,.
\end{equation}
The Hamiltonian $\bfH_\rmTD$ is $\beta$-Hermitian,
i.e., $\bfH_\rmTD = \beta \, \bfH^+_\rmTD \, \beta$.
Here, $\beta$ is the Dirac $\beta$ matrix which takes
the role of the $\gamma^5$ matrix in the Weyl representation
(see Appendix~\ref{appa}).
The $\beta$--Hermitian operator $S_\rmTD$ in this case reads as
\begin{equation}
S_\rmTD = -\ii\beta \, \frac{\beta\,\gamma^5\,m}\calE\Theta
=-\ii \, \gamma^5 \, \frac m\calE\Theta\,,
\qquad
S_\rmTD = \beta\,S_\rmTD^+\,\beta \,.
\end{equation}
The transformation
\begin{align}
U_\rmTD = \exp(\ii S_\rmTD) =
\cosh\left(\frac m\pp\Theta\right)
+\gamma^5\frac\pp\calE\sinh\left(\frac m\pp\Theta\right)
\end{align}
fulfills the identity
\begin{equation}
U^+_\rmTD \, \beta \, U_\rmTD 
= \exp(\ii S_\rmTD) \beta \exp(\ii S_\rmTD) 
= \beta \,,
\end{equation}
i.e., it is $\beta$-unitary.
It therefore conserves the $\calP\calT$-symmetric
scalar product $\left< \psi | \beta | \phi \right>$.
Choosing [redefining, see Eq.~\eqref{Theta1}] $\Theta$ so that
\begin{subequations}
\begin{align}
\cosh\left( 2 \frac{m \, \Theta}{| \vec p|} \right) = & \;
\frac{|\vec p|}{\sqrt{\vec p^{\,2} - m^2}} \,,
\\[0.133ex]
\sinh\left( 2 \frac{m \, \Theta}{| \vec p|} \right) = & \;
\frac{m}{\sqrt{\vec p^{\,2} - m^2}} \,,
\end{align}
\end{subequations}
one obtains
\begin{align}
\label{calHTD}
\calH_\rmTD = U_\rmTD \, \bfH_\rmTD \, U^{-1}_\rmTD = 
\beta\frac\calE\pp\sqrt{\vec p\,^2-m^2}\,.
\end{align}
This amounts to the exact ultrarelativistic decoupling
transformation of the free tachyonic Hamiltonian,
in the helicity (``Weyl'') basis which has 
been shown to lead to a very efficient description
of the tachyonic bispinor solutions~\cite{JeWu2012epjc,JeWu2013isrn,JeWu2014}.
The Taylor series expansion of $\calH_\rmTD$ 
gives rise to the terms
\begin{align}
\calH_\rmTD \approx & \;
\beta\left(\calE-\frac{m^2}{2\calE}
-\frac{m^4}{8\calE^3}\right) 
\nonumber\\[2.133ex]
=& \; -\beta \, \vec\Sigma \cdot \vec p \,
\left(1 - \frac{m^2}{2 | \vec p|^2} 
- \frac{m^4}{8 | \vec p|^4}\right)\,,
\end{align}
which are the correction terms in the high-energy limit.
For the tachyonic case, the eigenstates of the 
Hamiltonian~\eqref{calHTD} are still given by Eq.~\eqref{eigenS},
but one has to replace $E = \sqrt{\vec k^2 + m^2} \to 
\sqrt{\vec k^2 - m^2}$, in the sense of the 
tachyonic dispersion relation.

%
%
\section{Transformation With Gravitational Coupling}
\label{sec3}

%
%
\subsection{Gravitational Tardyonic Transformation}
\label{sec3A}

The study of the
gravitationally coupled Dirac equation, for massless 
particles,  was initiated by the question of how the
neutrinos (assumed by symmetry to be strictly massless in the Original Standard
Model) interact with gravitational fields~\cite{BrWh1957}.
We follow this route and start from the gravitationally coupled 
Dirac--Schwarzschild Hamiltonian~\cite{JeNo2013pra}
\begin{equation}
\label{HDS}
H_\rmDS = \frac12\left\{\vec \alpha \cdot \vec p, 1-\rr\right\}
+ \beta\,m\left(1-\rt\right)\,.
\end{equation}
After the initial transformation into the Weyl basis, one finds 
for $\bfH_\rmDS = U \,H_\rmDS \,U_1^{-1}$ [see Eq.~\eqref{defU}]
\begin{align}
\label{bfHDS}
\bfH_\rmDS =& \; 
\frac\beta2\left\{\calE,1-\rr\right\} + \gamma^5\,m\left(1-\rt\right)
\nonumber\\[2.133ex]
=& \left(\begin{array}{cc}
-\frac12\left\{\sdp,1-\rr\right\} &  m\left(1-\rt\right)\\
 m\left(1-\rt\right) & \frac12\left\{\sdp,1-\rr\right\}
\end{array}\right)\,.
\end{align}
In order to proceed with the ultrarelativistic decoupling
transformation, we identify the odd part $\calO_\rmDS$ 
of $\bfH_\rmDS$ and define
\begin{equation}
\label{known1}
\calO_\rmDS  = \gamma^5\,m\left(1-\rt\right)\,,\quad
S_\rmDS = -\ii\frac\beta4\left\{\calO_\rmDS, \frac1\calE\right\}\,,
\end{equation}
where $S_\rmDS$ is Hermitian.
The unitary transformation $U_\rmDS = \exp(\ii \, S_\rmDS)$
is applied to calculate
$\calH_\rmDS = U_\rmDS \, \bfH_\rmDS \, U_\rmDS^{-1}$,
perturbatively,
\begin{align}
\label{transform}
\bfH'_\rmDS \approx &  \;
\bfH_\rmDS + \frac{\ii^1}{1!} [ S_\rmDS, \bfH_\rmDS ]
+\frac{\ii^2}{2!} [S_\rmDS,[ S_\rmDS, \bfH_\rmDS  ]] + \dots
\end{align}
which is a series of nested commutators,
as with the classic Foldy--Wouthuysen transformation~\cite{FoWu1950}.
In the following, we carry the calculation
to first order in the Schwarzschild radius 
$r_s$ (first order in $G$) and keep 
inverse powers of $\calE$ up to order $1/\calE$.

It is advantageous to write the Hamiltonian~\eqref{bfHDS} as 
\begin{equation}
\bfH_\rmDS = \beta\,\calE-\frac\beta2\left\{\calE,\rr\right\} +
\calO_\rmDS \,.
\end{equation}
The first commutator is given as 
\begin{align}
\left[S_\rmDS, \bfH_\rmDS \right]
=&\; \left[S_\rmDS, \beta\,\calE\right]
-\left[S_\rmDS, \frac\beta2\left\{\calE,\rr\right\}\right]
\nonumber\\[1.133ex]
& \; +\left[S_\rmDS, \calO_\rmDS \right]\,.
\end{align}
Let us investigate the first commutator 
$\left[S_\rmDS, \beta\,\calE\right]$,
for which one finds after a somewhat tedious
calculation,
\begin{equation}
\label{COM1first}
[S_\rmDS, \beta\calE] = \ii\calO_\rmDS
+ \frac\ii4 \frac1\calE \,
[\calE, [ \calE, \calO_\rmDS ]] \, \frac1\calE \,,
\end{equation}
where the double commutator is proportional to a 
three-dimensional Dirac-$\delta$ function plus a spin orbit coupling term,
\begin{align}
[\calE, [ \calE, \calO_\rmDS ]] 
=& \; -2\pi\,r_s\,\delta^{(3)}(\vec r) 
- r_s \frac{\vec \Sigma \cdot \vec L}{r^3} \,,
\end{align}
which is of order unity in the expansion in inverse powers of $\calE$.
Despite the fact that the double commutator has two instances of
the operator $\calE$, the commutators ensure that these instances of
$\calE$ act {\em only} on $\calO_\rmDS$, and {\em not} on the 
reference state wave
function, which would otherwise generate inverse powers of 
$\calE$. For the
operator $\calE$ (or the inverse thereof) to be the ``dominant term'', it
must operate on a wave function describing a high-energy particle. Thus
\begin{align}
\frac1\calE\, [\calE, [\calE, \calO_\rmDS ]]\,\frac1\calE
=&\;\calO\left(\frac1{\calE^2}\right) \rightarrow 0\,.
\end{align}
Alternatively, one may observe that,
when using the Weyl free--spinors given in 
Eq.~\eqref{eigenS} as reference states, the 
the expectation values of both the Dirac--$\delta$ function and
the spin--orbit coupling term $(\vec\Sigma\cdot\vec L/r^3)$ vanish
for both diagonal as well as off-diagonal matrix elements.
In conclusion, to the order relevant for our investigation, we can replace
\begin{equation}
[S_\rmDS, \beta\,\calE] \; \to \; \ii\calO_\rmDS\,,
\end{equation}
in our approximation. This relation
ensures the odd terms will be canceled out 
to the first order in $\calO_\rmDS$
when calculating the transformed Hamiltonian $\calH_\rmDS$ according 
to Eq.~\eqref{transform}.  One also establishes that 
\begin{subequations}
\begin{align}
\left[S_\rmDS, \frac\beta2\left\{\calE, \rr\right\} \right]
=& \; \ii \gamma^5 m \rr \,,
\\[1.133ex]
\label{known2}
\left[S_\rmDS, \calO_\rmDS \right] =& \; \ii\,\beta\left(- \frac{m^2}\calE
+ \frac12m^2\left\{\frac1\calE,\rr\right\}\right)\,,
\end{align}
\end{subequations}
so that the first commutator becomes
\begin{equation}
\left[S_\rmDS, \bfH_\rmDS \right]
= \ii\left(\calO_\rmDS + \gamma^5 m \rr - \beta\frac{m^2}\calE
+ \frac{\beta\,m^2}{2}\left\{\frac1\calE,\rr\right\}\right).
\end{equation}
The double commutator is then 
\begin{align}
& \left[S_\rmDS,\left[S_\rmDS, \bfH_\rmDS \right]\right]
= \ii\left( \left[S_\rmDS,\calO_\rmDS \right]
+ \left[S_\rmDS,\gamma^5 m \rr\right]
\right.
\nonumber\\[1.133ex]
& \; \qquad \left.
- \left[S_\rmDS,\beta\frac{m^2}\calE\right]
+ \left[S,\frac{\beta\,m^2}{2} \, \left\{\frac1\calE,\rr\right\}\right]
\right)\,,
\end{align}
where the first term is known from Eq.~\eqref{known2}.
The other relevant commutators are
\begin{subequations}
\begin{align}
%
\left[S_\rmDS, \gamma^5 m \rr\right]
=&\, \ii \beta\frac{m^2}2\left\{\frac1\calE,\rr\right\} \,,
\\[1.133ex]
- \left[S_\rmDS, \beta\frac{m^2}\calE\right] = & \;
\calO\left( \frac{1}{\calE^2} \right) \to 0 \,,
\\[1.133ex]
\left[S_\rmDS, \frac12\,\beta\,m^2\left\{\frac1\calE,\rr\right\}\right] =& \;
\calO\left( \frac{1}{\calE^2} \right) \to 0 \,.
\end{align}
\end{subequations}
We then have
\begin{equation}
\left[S_\rmDS, \left[S, \bfH_\rmDS \right]\right]
= \beta\frac{m^2}\calE - \beta m^2\left\{\frac1\calE,\rr\right\} \,,
\end{equation}
where again we neglect higher-order terms.
Because $S_\rmDS$ carries an inverse power of $\calE$,
we can neglect the triple commutator,
\begin{equation}
\left[S_\rmDS, \left[S_\rmDS, 
\left[S_\rmDS, \bfH_{DS}\right]\right]\right]
=\calO\left(\frac1{\calE^2}\right)
\to 0 \,.
\end{equation}
Thus
\begin{align}
\bfH'_\rmDS =&\, \bfH_\rmDS 
+ \ii\, \left[S_\rmDS, \bfH_\rmDS \right]
+ \frac{\ii^2}{2!} \, 
\left[S_\rmDS, \left[S_\rmDS, \bfH_\rmDS \right]\right]
\nonumber\\[1.133ex]
=&\,\beta\left(\calE+\frac{m^2}{2\calE}
-\frac12\left\{\calE,\rr\right\}\right)
+ \calO'_\rmDS \,,
\end{align}
where
\begin{equation}
\calO'_\rmDS= - \gamma^5 \, m \, \rr \,.
\end{equation}
The second iteration of the transform with
\begin{equation}
S'_\rmDS = -\ii\frac\beta4\left\{\calO'_\rmDS, \frac1\calE\right\}\,,
\qquad U'_\rmDS = \exp(\ii S'_\rmDS) \,,
\end{equation}
will serve only to eliminate the remaining odd term. Thus,
the final result for $\calH_\rmDS = U'_\rmDS \, \bfH'_\rmDS \, 
U'^{-1}_\rmDS$
reads as
\begin{equation}
\label{calHDS}
\calH_\rmDS = \beta\left(\calE+\frac{m^2}{2\calE}
-\frac12\left\{\calE,\rr\right\}\right)
\end{equation}
for a gravitationally coupled high-energy Dirac particle.

%
%
\subsection{Gravitational Tachyonic Transformation}
\label{sec3B}

We start from the tachyonic, gravitationally coupled (TG) 
Dirac Hamiltonian derived in Appendix~\ref{appc},
\begin{equation}
\label{HTG}
H_{\rmTG} = \frac12\left\{\alphap,\left(1-\frac{r_s}r\right)\right\}
+ \beta\,\gamma^5\,m\,\left(1-\frac{r_s}{2r}\right)\,.
\end{equation}
The initial rotation into the Weyl basis of the Dirac algebra
using the transformation $U$ defined in Eq.~\eqref{defU} 
leads to the Hamiltonian $\bfH_\rmTG = U \, H_{\rmTG} \, U^{-1}$,
which reads as
\begin{align}
\bfH_\rmTG =& \; \frac{\beta}{2} \, \{\calE, 1 - \rr \} +
\beta \, \gamma^5 \, m\, \left( 1 - \rt \right)
\nonumber\\[0.133ex]
=& \; \left(\begin{array}{cc}
-\frac12\{\sdp, 1 - \rr \} & m\, \left( 1 - \rt \right) \\[1.133ex]
-m\, \left( 1 - \rt \right) & \frac12\{\sdp, 1 - \rr \}
\end{array}\right)\,,
\end{align}
where $\calE$ has been defined in Eq.~\eqref{defcalE}.
One identifies the odd part of the Hamiltonian $\bfH_\rmTG$
and writes
\begin{equation}
\label{S}
\calO_\rmTG = \beta \, \gamma^5 \,m\left(1-\rt\right)\,,\quad
S_\rmTG = -\ii \frac{\beta}{4} \,
\left\{\calO_\rmTG, \frac{1}{\calE} \right\}\,.
\end{equation}
The $\beta$-unitary transformation $U_\rmTG = \exp(\ii \, S_\rmTG)$ 
is applied to calculate $\bfH'_\rmTG = U_\rmTG \, \bfH_\rmTG \, U_\rmTG^{-1}$,
perturbatively,
\begin{equation}
\bfH'_\rmTG =
\bfH_\rmTG + \frac{\ii^1}{1!} [ S_\rmTG, \bfH_\rmTG ]
+ \frac{\ii^2}{2!}[S_\rmTG,[ S_\rmTG, \bfH_\rmTG ]] + \dots
\end{equation}
in full analogy with the Dirac--Schwarzschild Hamiltonian.
After a somewhat tedious calculation, neglecting 
(as before) the Dirac-$\delta$ and spin--orbit coupling terms, one finds
\begin{subequations}
\begin{align}
\label{TG1}
\left[S_\rmTG, \beta\calE\right] =& \; \ii\calO_\rmTG \,,
\\[1.133ex]
\label{TG2}
\left[S_\rmTG, \frac\beta2\left\{\calE,\rr\right\}\right] = & \;
\ii \ \beta \, \gamma^5 m \, \rr \,,
\\[1.133ex]
\label{TG3}
\left[S_\rmTG, \calO_\rmTG \right] =& \;
\ii\beta\,m^2\frac1\calE
-\ii\beta\, \frac{m^2}{2} \left\{\frac1\calE,\rr\right\} \, .
\end{align}
\end{subequations}
The first commutator becomes
\begin{align}
[S_\rmTG, \bfH_\rmTG] =& \;
\ii\left(\calO_\rmTG - \beta \gamma^5 m \rr +
\beta\frac{m^2}\calE \right.
\nonumber\\[1.133ex]
& \; \left. 
-\frac{\beta m^2}{2} \left\{\frac1\calE,\rr\right\}\right).
\end{align}
The double nested commutator is
\begin{align}
& [S_\rmTG,[S_\rmTG, H]] =
\ii\left([S_\rmTG, \calO_\rmTG ] -
\left[S_\rmTG, \beta\, \gamma^5 m \rr\right]
\right.
\nonumber\\[1.133ex]
& \; \left.  \qquad
+ \left[S_\rmTG, \beta\frac{m^2}{\calE} \right]
- \left[S_\rmTG, \beta\, \frac{m^2}{2} \,
\left\{\frac1\calE,\rr\right\}\right] \right) \,,
\end{align}
The last two commutators are of order $1/\calE^2$ 
and can therefore be neglected.
With the help of the result
\begin{equation}
\label{TG4}
\left[S,\beta \gamma^5 \, m \, \rr\right]
= \ii \frac{\beta m^2}{2} \left\{\frac1\calE,\rr\right\} \,.
\end{equation}
and with Eq.~\eqref{TG3}, one finds
\begin{equation}
[S_\rmTG,[S_\rmTG,\bfH_\rmTG]] = -\beta\,m^2\frac1\calE
+ \beta\, m^2 \left\{\frac1\calE,\rr\right\} \,.
\end{equation}
Thus,
\begin{align}
\bfH'_\rmTG =&\, \bfH_\rmTG 
+ \ii\, \left[S_\rmTG, \bfH_\rmTG \right]
+ \frac{\ii^2}{2!} \, \left[S_\rmTG,
\left[S_\rmTG, \bfH_\rmTG \right]\right]
\nonumber\\[1.133ex]
=&\,\beta\left(\calE+\frac{m^2}{2\calE}
-\frac12\left\{\calE,\rr\right\}\right)
+ \calO'_\rmTG\,,
\end{align}
where
\begin{equation}
\calO'_\rmTG= \beta \gamma^5 \, m \rr \,.
\end{equation}
A second transformation with 
\begin{equation}
S'_\rmTG = -\ii\frac\beta4\left\{\calO'_\rmTG, \frac1\calE\right\}\,,
\qquad U'_\rmTG = \exp(\ii S'_\rmTG) \,,
\end{equation}
eliminates $\calO'_\rmTG$, and we obtain the following
final result for $\calH_\rmTG = U'_\rmTG \, \bfH'_\rmTG \, {U'}^{-1}_\rmTG$,
\begin{equation}
\label{calHTG}
\calH_\rmTG = \beta\left(\calE-\frac{m^2}{2\calE}
-\left\{\frac{\calE}{2},\rr\right\} \right)\,.
\end{equation}
It differs from the result given in Eq.~\eqref{calHDS}
only in the sign of the kinetic term $-m^2/(2\calE)$,
due to the tachyonic dispersion relation.

%
%
\section{Conclusions}
\label{sec4}

We have studied the ultrarelativistic decoupling transformation
for the free Dirac equation (Sec.~\ref{sec2A}), 
and for the free tachyonic Dirac equation (Sec.~\ref{sec2B}).
These transformations lead to a full separation 
of the Dirac equation in the helicity basis.
Unlike the exact Foldy--Wouthuysen transformation,
which transforms the free Dirac Hamiltonian 
into the form $\beta \, \sqrt{\vec p^2 + m^2}$
(see Ref.~\cite{BjDr1964}), the ultrarelativistic transformation 
leads to a separation in the 
helicity basis, with the transformed Hamiltonian being 
proportional to $(-\beta \, \vec \Sigma \cdot \vec p)$
[see Eqs.~\eqref{calHFD} and~\eqref{calHTD}].
The eigenstates of this Hamiltonian are naturally 
obtained in the helicity basis
[see Eq.~\eqref{eigenS}] and are formally identical
(upon a redefinition of the energy parameter $E$)
to the eigenstates of the massless Dirac equation
(see Chap.~2.4.3 on p.~87 of Ref.~\cite{ItZu1980}).
The latter eigenstates are known to transform 
under the fundamental $(\tfrac12 \,, 0)$ representation
of the Lorentz group; the ``helicity of the massless spinors
does not flip upon a Lorentz transformation''.
This observation is intimately linked to the 
fact that massless Dirac spinors describe particles which 
always move at the speed of light;
it is impossible to ``overtake'' the particle, which 
otherwise leads to a helicity flip (see Ref.~\cite{JeWu2014}).

The initial unitary transformation $U$ given in Eq.~\eqref{defU}
transforms the Dirac equation into the Weyl basis
(see Appendix~\ref{appa}), which is naturally identified 
as the {\em ultrarelativistic basis} for the description of the 
Dirac algebra: Namely, the Dirac $\vec \alpha$ matrices 
are replace, in the Weyl basis, by matrices $(-\beta \, \vec \Sigma \cdot \vec p)$,
which are diagonal in the ($2 \times 2$)-spinor space
and describe the Hamiltonian for a massless Dirac particle.
The Dirac and Weyl representations of the Clifford algebra are
complementary: In the Dirac basis, the ``dominant term''
in the Hamiltonian is $\beta \, m$, and the odd (off-diagonal)
kinetic terms $\vec\alpha \cdot \vec p$ are eliminated 
by the Foldy--Wouthuysen transformation. In the Weyl basis,
the kinetic term $(-\beta \, \vec \Sigma \cdot \vec p)$ is
diagonal (``dominates in the ultrarelativistic limit), 
while the off-diagonal mass term $\gamma^5 \, m$ needs to 
be eliminated by the ultrarelativistic decoupling transformation.

With gravitational coupling in a central, static field, the dominant attractive
term is found to be described by the replacement $\calE \to
\left\{\frac{\calE}{2},1 - \rr\right\}$ in Eqs.~\eqref{calHDS}
and~\eqref{calHTG}, where $\calE$ is the energy operator defined in
Eq.~\eqref{defcalE}. This replacement holds both for tardyons and tachyons and
is a consequence of the structure of the Dirac--Schwarzschild Hamiltonian given
in Eqs.~\eqref{HDS} and~\eqref{HTG}.  Namely, the dominant interaction in the
high-energy limit is given by the anticommutator correction
$\frac12\left\{\alphap,\left(1-\frac{r_s}r\right)\right\}$ in the original
Hamiltonians (before ultrarelativistic decoupling) given in Eqs.~\eqref{HDS}
and~\eqref{HTG}.  The somewhat surprising observation that high-energy tachyons
are attracted by gravitational fields finds a natural explanation in the
energy-mass equivalence, and in the observation that both tachyons as well as
tardyons travel at speeds very close to the speed of light in the high-energy
limit.  Indeed, the only difference in the effective high-energy
Hamiltonians~\eqref{calHDS} and~\eqref{calHTG} lies in the sign of the kinetic
term $\pm m^2/(2 \calE)$, which is due to the changes in the dispersion
relation for tardyons as opposed to tachyons.
Higher-order corrections to the gravitational coupling
are discussed in Appendix~\ref{appd}.

The ultrarelativistic decoupling transformation should
find applications beyond the description of gravitational
interactions, for highly relativistic particles subject 
to electromagnetic fields, and further applications to 
``nearly massless'' electrons in graphene can be imagined
(here, the ``speed of light'' is replaced by the Fermi
velocity $v_F$, and dislocation potentials are added ``by hand'',
see Ref.~\cite{PeNiCN2007}).

\appendix

%
%
\section{Dirac and Weyl Basis}
\label{appa}

In the Dirac basis, we have
\begin{equation}
\vec\alpha = \left( \begin{array}{cc} 
0 & \vec\sigma \\
\vec\sigma & 0 
\end{array} \right) \,,
\qquad
\beta = \left( \begin{array}{cc}
\mathbbm{1}_{2 \times 2} & 0 \\
0 & -\mathbbm{1}_{2 \times 2} 
\end{array} \right) \,.
\end{equation}
The $\gamma$ matrices in the Dirac basis are 
$\gamma^0 = \beta$ and
\begin{equation}
\vec\gamma = \left( \begin{array}{cc}
0 & \vec\sigma \\
-\vec\sigma & 0
\end{array} \right) \,,
\qquad
\gamma^5 = \left( \begin{array}{cc}
0 & \mathbbm{1}_{2 \times 2}  \\
\mathbbm{1}_{2 \times 2} & 0
\end{array} \right) \,.
\end{equation}
In the Weyl basis, we have
\begin{equation}
\vec\alpha_\rmW = \left( \begin{array}{cc}
-\vec\sigma & 0 \\
0 & \vec\sigma 
\end{array} \right) \,,
\qquad
\beta_\rmW = \left( \begin{array}{cc}
0 & \mathbbm{1}_{2 \times 2}  \\
\mathbbm{1}_{2 \times 2} & 0
\end{array} \right) \,.
\end{equation}
We define the vector of $\vec\alpha_\rmW$ matrices
so that the ``upper'' solution includes the 
left-handed neutrino, whereas the ``lower'' spinor
contains the right-handed Dirac antineutrino 
[see Eq.~\eqref{eigenS}].
The $\gamma$ matrices in the Weyl basis are
$\gamma_\rmW^0 = \beta_\rmW$ and
\begin{equation}
\vec\gamma_\rmW = \left( \begin{array}{cc}
0 & \vec\sigma \\
-\vec\sigma & 0
\end{array} \right) \,,
\qquad
\gamma_\rmW^5 = \left( \begin{array}{cc}
-\mathbbm{1}_{2 \times 2} & 0  \\
0 & \mathbbm{1}_{2 \times 2} 
\end{array} \right) \,,
\end{equation}
so that $\gamma^5_\rmW = - \beta$.
We notice that 
$\vec\alpha_\rmW = \beta_\rmW \, \vec\gamma_\rmW$
and also 
$\vec\alpha_\rmW = -\beta \, \vec\Sigma$
where 
\begin{equation}
\vec\Sigma = 
\left( \begin{array}{cc}
\vec\sigma & 0 \\
0 & \vec\sigma 
\end{array} \right) \,.
\end{equation}
The vector of Pauli spin matrices is denoted as $\vec\sigma$.
Note that some authors define the vector $\vec\gamma_\rmW$
with the opposite sign, which also reverses the 
sign of $\gamma^5_\rmW = \ii \, 
\gamma^0_\rmW \, \gamma^1_\rmW 
\gamma^2_\rmW \, \gamma^3_\rmW $.

Incidentally, the Coulomb coupling identifies 
particles (which are attracted) and antiparticles (which are repulsed).
It is interesting to verify whether the interpretation is preserved
under the transformation to the Weyl representation. 
We start from the Hamiltonian
\begin{equation} 
H_C = \vec\alpha \cdot \vec p - \frac{Z\alpha}{r} \,,
\end{equation}
which describes a massless particles in a Coulomb field
(here, $Z$ is the nuclear charge number, while $\alpha$
is the fine-structure constant).
Transformation to the Weyl representation is accomplished
by the rotation
\begin{align}
\bfH_C =& \; U \, H_C \, U^{-1} = 
-\beta \, \vec\Sigma \cdot \vec p - \frac{Z\alpha}{r}
\nonumber\\[0.1133ex]
= & \;
\left( \begin{array}{cc} 
-\vec \sigma \cdot \vec p - \frac{Z\alpha}{r} & 0 \\
0 & \vec \sigma \cdot \vec p - \frac{Z\alpha}{r} \\
\end{array} \right) \,.
\end{align}
A comparison with Eq.~\eqref{eigenS} reveals that 
states with positive unperturbed energy (positive eigenvalue of the 
operator $-\vec \sigma \cdot \vec p$ for the upper spinor
and positive eigenvalue of $\vec \sigma \cdot \vec p$ for the 
lower spinor) are attracted by the Coulomb field.
By contrast, states with negative unperturbed energy (negative eigenvalue of the
operator $-\vec \sigma\cdot \vec p$ for the upper spinor
and negative eigenvalue of $\vec \sigma\cdot \vec p$ for the 
lower spinor) are repulsed by the Coulomb field.

%
%
\section{Operators}
\label{appb}

We wish to explore the application of the 
operator $1/\calE = -1/(\vec \Sigma \cdot p)$ 
to a reference state wave function.
To this end, we assume that $f = f(\vec r)$ is a test function,
and we defined the Fourier transform $\calF$ and 
Fourier backtransform $\calF^{-1}$ as follows,
\begin{align}
(\calF f)(\vec k) =& \;
\int \dd^3 r \exp(-\ii \vec k \cdot \vec r) \, f(\vec r) \,,
\\[0.1133ex]
(\calF^{-1} F)(\vec r) =& \; \int \frac{\dd^3 k}{(2 \pi)^3} \,
\exp(\ii \vec k \cdot \vec r) \, f(\vec k) \,.
\end{align}
One first multiplies the operator 
\begin{equation}
1/\calE \to -1/(\vec\Sigma \cdot \vec k) =
-\frac{\vec\Sigma \cdot \vec k}{\vec k^2}
\end{equation}
in Fourier space and then transforms 
back to coordinate space,
\begin{align}
\left(\frac{1}{\calE} f\right)(\vec r) =& \;
\left[ \calF^{-1} \left( -\frac{\vec\Sigma \cdot \vec k}{\vec k^2} 
\left(\calF \, f \right)(\vec k) \right) \right](\vec r)
\nonumber\\[0.1133ex]
=& \; - \int \frac{\dd^3 k}{(2 \pi)^3} \,
\int\dd^3 r' \,
\frac{\vec \Sigma \cdot \vec k}{\vec k^2}
\ee^{\ii \vec k \cdot \left(\vec r - \vec r'\right)} \, 
f(\vec r') \,.
\end{align}
For a reference state with a special value $\vec k_s$
of the wave vector [see Eq.~\eqref{eigenS}],
\begin{equation}
f_s(\vec r) = \left. \psi_1(\vec r) \right|_{\vec k \to \vec k_s, t = 0} 
= \left( \begin{array}{c} a_-(\vec k_s) \\ 0 \end{array} \right) \, 
\exp\left(\ii \vec k_s \cdot \vec r \right) \,,
\end{equation}
one has
\begin{equation}
(\calF \, f_s)(\vec k) = 
\left( \begin{array}{c} a_-(\vec k_s) \\ 0 \end{array} \right) \, 
\delta^{(3)}(\vec k - \vec k_s)
\end{equation}
and so
\begin{equation}
-\frac{\vec \Sigma \cdot \vec k}{\vec k^2}
(\calF \, f_s)(\vec k) = 
\frac{1}{|\vec k_s| }
\left( \begin{array}{c} a_-(\vec k_s) \\ 0 \end{array} \right) \, 
\delta^{(3)}(\vec k - \vec k_s) \,,
\end{equation}
whose Fourier backtransform is 
\begin{equation}
\left( \frac{1}{\calE} f_s\right)(\vec r) =
\frac{1}{|\vec k_s|} \, 
\left( \begin{array}{c} a_-(\vec k_s) \\ 0 \end{array} \right) \, 
\exp\left( \ii \vec k_s \cdot \vec r \right) \,.
\end{equation}
This corresponds to the naive result that we obtain when
interpreting the $\calE$ operator as an energy operator
and applying it to the eigenstates of the free 
Hamiltonian, given in Eq.~\eqref{eigenS}.

%
%
\section{Formalism for Gravitational Coupling}
\label{appc}

We here follow the conventions
used in Refs.~\cite{JeNo2013pra,Je2013}
for the flat--space and curved--space Dirac gamma matrices.
Specifically, the flat--space and curved--space Dirac gamma
matrices are distinguished in this Appendix
using a tilde ($\widetilde \gamma$) and an overline 
($\overline\gamma$) respectively. 
We draw inspiration from the book~\cite{Bo2011}
and denote indices related 
to a local Lorentz frame (``anholonomic basis'')
with capital Latin indices $A,B,C,\ldots = 0,1,2,3$.
The curved-space Dirac gamma matrices 
$\overline\gamma^\mu$ satisfy the condition that
\begin{equation}
\left\{\overline \gamma^\mu(x), \overline\gamma^\nu(x)\right\} =
2 \overline g^{\mu\nu}(x)\,,
\end{equation}
where $\overline g^{\mu\nu}(x)$ is the curved-space-time metric.
The $\overline\gamma^\mu(x)$ are expressed in 
terms of the flat-space Dirac $\widetilde\gamma$ matrices
$\widetilde\gamma^A$ as follows,
\begin{equation}
\overline\gamma^\mu(x) = e^\mu_A \, \widetilde\gamma^A \,, 
\qquad
\overline\gamma_\mu(x) = e_\mu^A \, \widetilde\gamma_A \,,
\end{equation}
where the $e^\mu_A$ are the coefficients which 
relate the locally flat Lorentz frame to the 
global space-time coordinates (the ``vierbein'').
Greek indices $\mu,\nu,\rho, \ldots=0,1,2,3$
denote the global coordinates.
Latin indices starting with $i,j,k, \ldots = 1,2,3,\ldots$
are reserved for ``spatial'' global coordinates,
which leaves $I,J,K, \dots = 1,2,3,\ldots$
for spatial coordinates in the anholonomic basis.
This notation addresses some ambiguities which could 
otherwise result from other 
approaches~\cite{Iv1969a,Iv1969b,BrWh1957,Bo1975prd,%
SoMuGr1977,Go1985,Ye2011,ZaMB2012}.
For example, unless the Dirac matrices 
are distinguished by overlining or 
tildes, the expression $\gamma^1$ could be associated with a
flat-space matrix $\widetilde\gamma^{I=1}$ or with a 
curved-space matrix $\overline\gamma^{i=1}$.
We use the ``West--Coast'' convention
for the flat-space metric, which we denote as
$\eta^{AB}= \eta_{AB} = \mbox{diag}(1,-1,-1,-1)$. 
The curved-space metric is recovered as
\begin{align}
\eta_{AB} =& \; 
\frac12\{\widetilde\gamma_A,\widetilde\gamma_B\} \,,
\\[1.133ex]
\overline g_{\mu\nu}(x) =& \; \frac12\{\overline\gamma_\mu(x),
\overline\gamma_\nu(x)\}
= e_\mu^A \, e_\nu^B \, \eta_{AB} \,,
\\[1.133ex]
\overline g^{\mu\nu}(x) =& \; \frac12\{\overline\gamma^\mu(x),
\overline\gamma^\nu(x)\}
= e^\mu_A \, e^\nu_B \, \eta^{AB} \,.
\end{align}
For the curved--space metric around a gravitational center,
we use the isotropic Schwarzschild metric in the Eddington
reparameterization~\cite{Ed1924}, i.e.
\begin{align}
\overline g_{\mu\nu} =& \; 
\operatorname{diag}(w^2,-v^2,-v^2,-v^2)\,,
\\[1.113ex]
\overline g^{\mu\nu} =& \; 
\operatorname{diag}(w^{-2},-v^{-2},-v^{-2},-v^{-2})\,,
\\[1.113ex]
\label{defwv}
w=& \; \frac{1 - \dfrac{r_s}{4r}}{1 + \dfrac{r_s}{4r}}\,,
\qquad
v = \left(1+\frac{r_s}{4r}\right)^2\,.
\end{align}
For the Schwarzschild geometry,
the vierbein coefficients read as follows,
\begin{align}
\label{coeff}
e^0_\mu =& \; \delta^0_\mu\,w\,,\qquad
e^A_\mu = \delta^A_\mu \,v\,,
\\[1.133ex]
e^\mu_0 =& \; \frac{\delta^\mu_0}{ w }\,,\qquad
e^\mu_A =\frac{\delta^\mu_A}{v}\,.
\end{align}
Here, $\delta^\mu_A = \delta^A_\mu$ denotes the Kronecker-$\delta$
(which is of course equal to unity 
for the two indices being equal and zero otherwise). 

In full analogy with the case of a ``normal'' massive
Dirac particle (see Refs.~\cite{Je2013,JeNo2013pra}),
we write the Dirac action for a tachyon in curved spacetime as
\begin{align}
\label{accion}
S =& \; \int d^4x \; \sqrt{-\det \overline g(x)} 
\\[0.133ex]
& \; \times \overline\psi(x) \, 
\overline\gamma^5(x) \,
\left(\frac\ii2\gamma^\rho(x)
\overleftrightarrow\nabla_\rho -
\overline\gamma^5(x) \, m\right)\psi(x)\,,
\nonumber
\end{align}
where $\overline\psi(x) \, \overline\gamma^5(x) $
takes the role of the ``tachyonic adjoint'' 
(see Ref.~\cite{JeWu2013isrn}) and 
\begin{align}
\nabla_\rho =& \; \partial_\rho-\Gamma_\rho(x)\,,
\\[0.133ex]
\Gamma_\mu(x) =& \; \frac{\ii}{4} \,
\omega_\mu^{AB}(x) \, \widetilde\sigma_{AB}  \,,
\qquad
\widetilde\sigma_{AB} = \frac{\ii}{2} \, 
\left[ \widetilde\gamma_A,\widetilde\gamma_B \right]\,,
\\[0.133ex]
\omega_\nu^{A B}(x)
=& \; e^A_\mu \, \nabla_\nu \, e^{\mu B}
= e^A_\mu \, \partial_\nu \, e^{\mu B} +
e^A_\mu \, \Gamma^\mu_{\nu \lambda} \, e^{\lambda B} \,.
\end{align}
Here, the $\Gamma^\mu_{\nu \lambda}$ are the 
Christoffel symbols, and the $\omega_\nu^{A B}(x)$
are known as the Ricci rotation coefficients.
Under a spinor Lorentz transformation
with generators $\Omega^{AB}(x)$,
\begin{equation}
\psi'(x') = S(\Lambda(x)) \,\psi(x) =
\exp\left( - \frac{\ii}{4} \, \Omega^{AB}(x) \,  
\widetilde\sigma_{AB}  \right) \,
\psi(x)
\end{equation}
we have covariance, i.e.,
$\nabla'_\nu \, \psi'(x) =
\nabla'_\nu \, [ S(\Lambda(x)) \psi(x) ] =
(\partial_\nu - \Gamma'_\mu) \, [ S(\Lambda(x)) \psi(x) ] =
S(\Lambda(x)) \, \nabla_\nu \, \psi(x)$,
where the tranformed Ricci rotation 
coefficients $\Gamma'_\mu = \frac{\ii}{4} \,
{\omega'}_\mu^{AB}(x) \, \widetilde\sigma_{AB}$
are calculated with respect to the transformed local coordinates.

The curved-space fifth current 
$\overline\gamma^5(x)$ needs to be clarified.
Adopting Eq.~(18) of Ref.~\cite{BrWh1957} for West--Coast sign conventions,
one finds
\begin{equation}
\widetilde\gamma^5(x) = \frac\ii{4!}
\frac{\epsilon_{\mu\nu\rho\delta}}{\sqrt{-\det g(x)}}
\widetilde\gamma^\mu(x)\,
\widetilde\gamma^\nu(x)\,
\widetilde\gamma^\rho(x)\,
\widetilde\gamma^\delta(x)\,,
\end{equation}
where $\epsilon$ is the fully antisymmetric 
Levi-Civita tensor, with $\epsilon_{0123} =1$.
We recall that the flat-space $\widetilde\gamma^5$ is 
\begin{equation}
\gamma^5 = \gamma_5 = \frac\ii{4!} \, \epsilon_{ABCD}
\widetilde\gamma^A \, 
\widetilde\gamma^B \, 
\widetilde\gamma^C \, 
\widetilde\gamma^D \,.
\end{equation}
For the Schwarzschild geometry, one easily finds
%
\begin{align}
\det \overline g(x) =&\,-w^2v^6\,,
\qquad
\sqrt{-\det \overline g(x)} = w\,v^3\,,
\\[1.113ex]
%
\overline\gamma^5(x) =& \; 
\overline\gamma_5(x) = 
\ii \widetilde\gamma_0\widetilde\gamma_1\widetilde\gamma_2\widetilde\gamma_3
= \widetilde\gamma^5 \equiv \gamma^5 \,,
\end{align}
%
i.e., the flat-- and curved--space $\gamma^5$ matrices are identical. 
Variation of the action~\eqref{accion} gives us
\begin{equation}
\left(\ii \overline\gamma^\mu \, \nabla_\mu-
\gamma^5 \, m\right)\psi=0\,,
\end{equation}
which can be rewritten as
\begin{equation}
\label{Ham1}
\ii(\overline\gamma^0)^2 \,
\partial_0 \psi = 
\left(
\overline\gamma^0 \, \overline\gamma^i\, p^i 
+ \ii\overline\gamma^0 \, \overline\gamma^\mu \, 
\Gamma_\mu + \gamma^5 m \right) \, \psi\,.
\end{equation}
An explicit calculation of the 
Ricci rotation coefficients show that 
\begin{equation}
\overline\gamma^0 \,
\overline\gamma^\mu \, \Gamma_\mu = 
- \frac{\vec\alpha\cdot\vec\nabla w}{2w^2v}
- \frac{\vec\alpha\cdot\vec\nabla v}{wv^2}\,,
\end{equation}
which is in agreement with Ref.~\cite{JeNo2013pra,Je2013},
where the $\vec\alpha = \widetilde\gamma^0 \, 
\vec{\widetilde\gamma}$ 
matrices are flat--space matrices.
With the help of Eq.~\eqref{Ham1}, one then finds that
$\ii\partial_t\psi = H\psi$, where
\begin{align}
H=\frac wv\alphap
+ \frac{\vec\alpha\cdot [\vec p,w]}{2v} 
+ \frac{w\vec\alpha\cdot [\vec p,v])}{v^2}
+ \beta\gamma^5 m\,w\,,
\end{align}
and $\beta=\widetilde\gamma^0$.
We now stretch the spatial coordinates with the 
help of the operator $v^{3/2}$, in analogy to the 
tardyonic case~\cite{JeNo2013pra,Je2013},
and find the $\gamma^5$--Hermitian Hamiltonian
\begin{equation}
\label{scale}
H' = v^{3/2}\,H\,v^{-3/2}
= \frac12
\left\{ \alphap, \frac{w}{v} \right\}+\beta \gamma^5\,m\,w\,,
\end{equation}
with $H' = \gamma^5 \, H'^+ \, \gamma^5$.
Approximating $w$ and $v$, according to Eq.~\eqref{defwv},
to the first order in gravity,
\begin{equation}
\label{approxwv}
w \approx 1-\frac{r_s}{2r}\,,\quad
v \approx 1+\frac{r_s}{2r}\,,
\end{equation}
one finds the tachyonic gravitationally (TG) coupled Hamiltonian
\begin{equation}
\label{initialHTG}
H_{\rm{TG}} = \frac12\left\{\alphap,\left(1-\frac{r_s}r\right)\right\}
+ \beta\,\gamma^5\,m\,\left(1-\frac{r_s}{2r}\right)\,.
\end{equation}
In the main body of the article,
standard notation is exclusively used for the 
{\em flat}-space Dirac matrices
(no overlining and no tildes), i.e., we denote 
the $\widetilde\gamma^\mu$ as $\gamma^\mu$.

%
%
\section{Higher-Order Terms}
\label{appd}

As discussed in Secs.~\ref{sec2} and~\ref{sec3}, the only difference between the effective
high--energy Hamiltonians for tardyons
[Eq.~\eqref{calHDS}] and
tachyons [Eq.~\eqref{calHTG}], derived in the main body of this work, is due the
different dispersion relations for (free) tardyons and tachyons,
while the gravitational interaction terms are identical 
to first order in $r_s$ and first order in $1/\calE$. 
This observation can be traced to the fact that the 
terms multiplying the kinetic operator
and the mass in Eqs.~\eqref{HDS} and~\eqref{HTG},
namely,
$X = 1 - \frac{r_s}{r}$, and
$Y = 1 - \frac{r_s}{2 r}$
fulfill the relationship
$Y^2/X = 1 + \calO(r_s^2)$.
One then easily reveals the cancellation mechanism 
for the terms of first order in $r_s$
by treating $X$ and $Y$ in the non-transformed 
Hamiltonians~\eqref{HDS} and~\eqref{HTG}
as constants.  However, this does not imply 
that gravitational effects are the same for tardyons and tachyons,
in higher orders of $G$ (higher orders of $r_s$).

The ultrarelativistic decoupling transformation,
keeping terms second order in $r_s$, 
and to the first order in $1/\calE$ (see Appendix~\ref{appb}),
is expected to lead to differences in the gravitational interaction terms.
In the calculation,
one needs to take into account the fact that 
in higher orders of the gravitational coupling constant, we cannot use the starting
Hamiltonians as defined in Eqs.~\eqref{HDS} and~\eqref{HTG}.
Instead, we must must use higher order
approximations to the gravitational terms,
which are otherwise neglected in Eq.~\eqref{approxwv}
(see also Refs.~\cite{Je2013,JeNo2013pra,JeNo2014jpa}).
These lead to the initial Hamiltonians
\begin{align}
H_{\rm ds}=& \; \frac12\left\{\alphap,1-\rr+\frac{9\,r_s^{2}}{16r^2}\right\}
\nonumber\\
&\;+\beta\,m\left(1-\rt+\frac{r_s^{\,2}}{8r^2}\right)
\end{align}
for tardyons and 
\begin{align}
H_{\rm td}=&\;\frac12\left\{\alphap,1-\rr+\frac{9\,r_s^{2}}{16r^2}\right\}
\nonumber\\
&\;+\beta\gamma^5\,m\left(1-\rt+\frac{r_s^{\,2}}{8r^2}\right)
\end{align}
for tachyons.
We then transform these Hamiltonians into the Weyl basis using the transform
$U$ defined in Eq.~\eqref{defU}. 
Calculations become tedious and lengthy.
One observation in generalizing the decoupling transformation
is that given a function $f=f(\vec r)$, then to first
order of $1/\calE$ one finds
\begin{equation}
\frac1\calE\,f\,\calE+\calE\,f\,\frac1\calE
=2\,f+\frac1\calE\left[\calE,\left[\calE,f\right]\right]\frac1\calE
\to 2\,f\,.
\end{equation}
As discussed in Sec.~\ref{sec3A}, this is due to the fact that the two
operators $\calE$ act only on the function $f(\vec r)$, and not on the
wavefunction, thus they do {\em not} give ``dominating'' energy terms.
After three iterations of the transform (per Hamiltonian), one
finds for tardyons
\begin{align}
\calH_{\rm ds}=&\;\beta\left( \calE + \frac{m^2}{2\calE}
-\frac12\left\{\calE,\rr\right\}+\frac9{32}\left\{\calE,\rs\right\}
\right.\nonumber\\&\;\left.
-\frac{7\,m^2}{64}\left\{\frac1\calE,\rs\right\}
+\frac{3\,m^2}{16}\rr\frac1\calE\rr\right)\,,
\end{align}
while for tachyons
\begin{align}
\calH_{\rm tg}=&\;\beta\left( \calE - \frac{m^2}{2\calE}
-\frac12\left\{\calE,\rr\right\}+\frac9{32}\left\{\calE,\rs\right\}
\right.\nonumber\\&\;\left.
+\frac{7\,m^2}{64}\left\{\frac1\calE,\rs\right\}
-\frac{3\,m^2}{16}\rr\frac1\calE\rr\right)\,.
\end{align}
The final two terms in these Hamiltonians have opposite signs,
indicating a difference in the gravitational interaction for tachyons and
tardyons.

\end{document}